\documentclass[showpacs,aps,graphicx,amsmath]{revtex4}
\usepackage{graphicx}
\usepackage{amsmath}
\usepackage{amssymb}

\begin{document}

\title{Scalable quantum computing based on stationary  spin qubits in coupled quantum dots inside double-sided optical microcavities\footnote{Published in  Sci. Rep. \textbf{4}, 7551  (2014).}}

\author{Hai-Rui Wei and  Fu-Guo Deng\footnote{Correspondence and
requests for materials should be addressed to F. -G. Deng
(fgdeng@bnu.edu.cn).}}

\address{Department of Physics, Applied Optics Beijing Area Major Laboratory, Beijing Normal University, Beijing 100875, China}

\date{\today }

\begin{abstract}
Quantum logic gates are the key elements in quantum computing. Here
we investigate the  possibility  of  achieving  a  scalable  and
compact quantum computing based on stationary electron-spin qubits,
by using the giant optical circular birefringence induced by
quantum-dot spins in double-sided optical microcavities as a result
of cavity quantum electrodynamics. We design the compact quantum
circuits for implementing universal and deterministic quantum gates
for electron-spin systems, including the two-qubit CNOT gate and the
three-qubit Toffoli gate. They are compact and economic, and they
do not require additional electron-spin qubits. Moreover, our
devices have good scalability and are attractive as they both are
based on solid-state quantum systems and the qubits are  stationary.
They are feasible with the current experimental technology, and both
high fidelity and high efficiency can be achieved when the ratio of the side leakage to the cavity decay is low.
\end{abstract}

\maketitle

Quantum logic gates are the basic elements to realize quantum
computation and quantum information processing. It is well known
that the controlled-not (CNOT) gate is one of the most efficient
quantum gates. CNOT gates supplemented with single-qubit rotations
are widely adopted as the standard model of universal quantum
computation \cite{uni,uni1,uni2,uni3,uni4}. It has been shown that
the synthesis of a general two-qubit gate requires 3 CNOT gates and
15 elementary one-qubit rotations in the worst case
\cite{3CNOT1,3CNOT2,3CNOT3}. The ``small-circuit" structure for
two-qubit gates in terms of  CNOT gates has been well solved
\cite{small}. In experiment, a single-qubit gate is easily
implemented by a local Hamiltonian or an external field, while the
two-qubit operations highly depend on the physical systems and there
are more demanding and imperfection to implement them. That is, it
is interesting to investigate the implementation of the two-qubit
CNOT gate in specific physical systems.

The implementation of multi-qubit gates is an important milestone
for a scalable quantum computing. However, the realization of a
generic multi-qubit gate is quite worse in terms of CNOT gates and
single-qubit rotations, and  the  implementation of a two-qubit gate
in a multi-qubit system is  usually more complex than that in a
two-qubit system. In 2004, Shende  \emph{et al.} \cite{3CNOT3} gave
the theoretical lower bound for multi-qubit gates, $[(4^n-3n-1)/4]$,
in terms of CNOT gates. However, up to now, the ``small-circuit"
structure and  the specific synthesis of the logic gates for
multi-qubit systems are two open questions. Among the three-qubit
gates, many efforts have been made in studying the fundamental
Toffoli gate which is not only a universal gate for classical
computing but also for quantum computing \cite{book,Toffoli}.
Together with Hadamard gates, Toffoli gates form a universal set of
quantum gates for quantum computation \cite{Toffoli}. Moreover,
Toffoli gate is a central building block in some quantum
cryptography protocols, phase estimation, and some quantum
algorithms. The optimal CNOT-gate cost of a Toffoli gate
\cite{Toffolicost} is 6. Fewer resources and simpler quantum
circuits are desired for an efficient quantum computation. It is
thus desirable to seek simpler schemes to directly implement Toffoli
gate.

Although many interesting protocols have been proposed to construct
universal quantum  gates,  it is still a big challenge to implement
quantum gates in experiment. The ones based on solid-state quantum
systems are especially attractive  because of their good scalability
and stability.  Quantum dot \cite{SWAP} (QD) is one of the promising
candidates for a solid-state qubit, due to the modern semiconductor
technology and the microfabrication technology. The long
electron-spin decoherence time ($T_2\sim\mu$s) of a  QD by using
spin-echo techniques \cite{cohertime1,cohertime2,cohertime3},
nanoscale confinement of electrons \cite{nanoscale1,nanoscale2}, the
preparation of the QD-spin superposition state
\cite{superposition1,superposition2}, the QD-spin detection
techniques \cite{spin-detection}, and the electron-spin manipulation
using picosecond/femtosecond optical pulses
\cite{spin-manipulate1,spin-manipulate2,spin-manipulate3,spin-manipulate4}
make an electron spin in a QD  an excellent candidate for the qubit
in solid-state quantum computation. In 2008, Hu \emph{et al.}
\cite{Hu1,Hu2} proposed a device, an excess electron confined in a
self-assembled In(Ga)As QD or a GaAs interface QD placed inside a
single-sided or a double-sided optical resonant cavity.  Many
interesting  tasks have been carried out on this quantum system
\cite{Hu3,Hu4,Hu5,Hu6,Appli1,weicnot,RenLPL,RenSR,WeiOEphoton,Appli2,repeater,Ren,wangtjhbsa}.
For example, utilizing this system, Hu \emph{et al.} \cite{Hu3,Hu4}
built a controlled-phase gate with a polarization photon  as the
control qubit and an electron spin as the target qubit. Bonato
\emph{et al.} \cite{Appli1} constructed a hybrid CNOT gate with an
electron spin as the control qubit and a polarization photon as the
target qubit. Also, a hybrid CNOT gate with a polarization photon as
the control qubit and an electron spin as the target qubit was
proposed recently \cite{weicnot}. In 2013, Ren \emph{et al.}
\cite{RenLPL} proposed a scheme for a photonic spatial-polarization
hyper-controlled-not gate assisted by a QD inside a one-sided
optical microcavity. In 2014, Ren and Deng \cite{RenSR} presented a
scheme for the hyper-parallel photonic quantum computation with
coupled QDs. The scheme for the CNOT gate on two photonic qubits
\cite{WeiOEphoton} was presented in 2013. A scheme \cite{Appli2} for
entanglement purification and concentration of electron-spin
entangled states and a quantum repeater scheme \cite{repeater} based
on QD spins in optical microcavities were introduced in 2011 and
2012, respectively.

In this paper, we investigate the possibility to achieve a compact
and scalable quantum computing based on stationary electron-spin
qubits. We construct two important universal quantum gates on
electron-spin systems, including the two-qubit CNOT gate and the
three-qubit Toffoli gate, by using the giant optical circular
birefringence induced by the electron spins in QDs confined in
double-sided optical microcavities as a result of cavity quantum
electrodynamics (QED). We give the compact quantum circuits and the
detailed processes for implementing these universal quantum gates.
The qubits of our gates are encoded on two orthogonal electron-spin
states of the excess electrons confined in QDs inside optical
resonant microcavities, denoted by $|\uparrow\rangle$ and
$|\downarrow\rangle$. A polarized single photon, denoted by
$|R\rangle$ or $|L\rangle$, plays a medium role. After the
input-output process of the single photon, the measurement on the
polarization of the output photon and some proper single-qubit
operations are performed on the electron-spin qubits, the evolutions
of these universal quantum gates are accomplished with the
probability of 100\% in principle. Our protocols have some features.
First, our quantum circuits for the universal quantum gates are
compact and economic, and they reduce the resources needed and the
errors as they do not require additional electron-spin qubits, just
a flying photon. Second, the double-sided QD-cavity system easily
reaches a large phase difference ($\pi$) between the uncoupled
cavity and the coupled cavity \cite{Hu2}, while it is a hard work in
a single-sided QD-cavity system.  Third, our gates allow for a
scalable and stable quantum computing as the qubits for the gates
are confined in solid-state quantum systems. Fourth, our schemes
work in a deterministic way if the photon loss caused by the optical
elements (such as  half-wave plates, and polarizing beam splitters)
and the detection inefficiency are negligible. Fifth, our schemes
are feasible with current technology. Both a high fidelity and a
high efficiency for each gate can be achieved when the ratio of the
side
leakage to the cavity decay is low.\\

\bigskip

{\large \textbf{Results}}

\begin{figure}[htbp]             
\centering\includegraphics[width=10 cm]{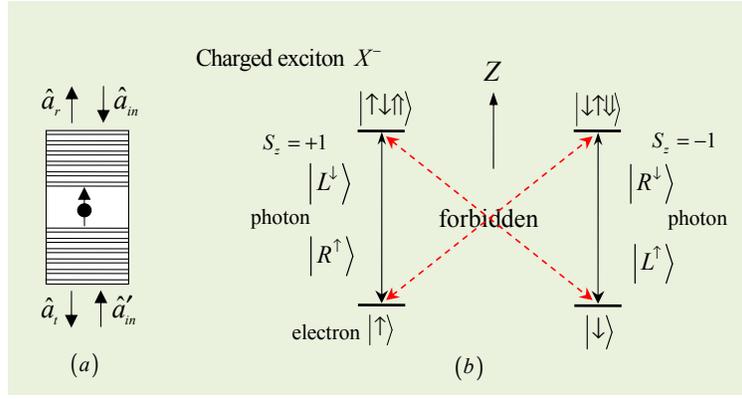} \caption{  (a)
Structure of a singly charged QD inside a double-sided optical
microcavity with circular cross-section. (b) Energy-level scheme of
a singly charged QD inside a double-sided optical microcavity with
the polarization allowed transition rules for the coupling photons.
$\vert R\rangle$ ($\vert L\rangle$) represents a right-circularly
(left-circularly) polarized photon. \label{figure1}}
\end{figure}

\textbf{Giant optical circular birefringence.} We exploit the
optical property \cite{Hu1,Hu2,Hu3,Hu4} of the QD-cavity system to
complete our schemes for implementing the CNOT gate and the Toffoli
gate on electron spins. The schematic diagram for a QD-cavity
system, a singly charged In(Ga)As QD or a GaAs interface QD placed
at the antinode of a resonant double-sided optical microcavity with
two symmetric and low loss partially reflective mirrors in the top
and the bottom \cite{Hu2}, employed in our protocols, is shown in
Fig. \ref{figure1}(a).  The negatively charged exciton (trion,
$X^-$), which consists of two electrons and one heavy hole
\cite{trion}, is the fundamental optical excitation, and it is
essential for optical transitions in a QD-cavity system. There are
two kinds of spin-dependent optical transitions, shown in Fig.
\ref{figure1}(b). The photon in the state $|R^\uparrow\rangle$ or
$|L^\downarrow\rangle$ ($s_z=+1$) couples to the dipole for the
transition from $|\uparrow\rangle$ to
$|\uparrow\downarrow\Uparrow\rangle$, and the photon  in the state
$|R^\downarrow\rangle$ or $|L^\uparrow\rangle$ ($s_z=-1$) couples to
the dipole for the transition from $|\downarrow\rangle$ to
$|\downarrow\uparrow\Downarrow\rangle$. Here the superscript arrows
of $|R\rangle$ ($|L\rangle$) indicate their propagation direction of
the photon along the normal direction of the cavity $z$ axis and the
circular polarization of the photons are dependent of their
propagation direction. $|\Uparrow\rangle$  and $|\Downarrow\rangle$
represent the two heavy-hole spin states with the $z$-direction spin
projections $|+\frac{3}{2}\rangle$ and $|-\frac{3}{2}\rangle$,
respectively. $|\uparrow\rangle$ and $|\downarrow\rangle$ represent
the electron-spin states with the $z$-direction spin projections
$|+\frac{1}{2}\rangle$ and $|-\frac{1}{2}\rangle$, respectively.

The change of the input photon state in the QD-cavity system can be
obtained by solving the Heisenberg equations of motion for a
QD-cavity system \cite{Heisenberg}
\begin{eqnarray}          \label{eq2}
\begin{split}
&\frac{d\hat{a}}{dt}=-\left[i(\omega_c-\omega)+\kappa+\frac{\kappa_s}{2}\right]\hat{a}
-\text{g}\sigma_{-}-\sqrt{\kappa}\,\hat{a}_{in} - \sqrt{\kappa}\,\hat{a}_{in}'+\hat{H},\\
&\frac{d\sigma_-}{dt}=-\left[i(\omega_{X^-}-\omega)+\frac{\gamma}{2}\right]\sigma_{-}-\text{g}\sigma_z\hat{a}+\hat{G},
\end{split}
\end{eqnarray}
and the input-output relation for the cavity
\begin{eqnarray}        \label{eq3}
\hat{a}_r = \hat{a}_{in}+ \sqrt{\kappa}\,\hat{a},\qquad\quad
\hat{a}_t=\hat{a}_{in}'+\sqrt{\kappa}\,\hat{a}.
\end{eqnarray}
The specific expression of the reflection and the transmission
coefficients of a realistic QD-cavity system is \cite{Hu2}:
\begin{eqnarray}       \label{eq4}
r(\omega)=1+t(\omega),\qquad
t(\omega)=\frac{-\kappa[i(\omega_{X^{-}}-\omega)+\frac{\gamma}{2}]}
           {[i(\omega_{X^{-}}-\omega)+\frac{\gamma}{2}][i(\omega_c-\omega)+\kappa+\frac{\kappa_s}{2}]+\text{g}^2}.
\end{eqnarray}
Here,  $\omega_c$, $\omega$, and  $\omega_{X^-}$ are the
frequencies of the cavity, the input photon, and the $X^-$
transition, respectively. $\hat{a}$ and $\sigma_-$ are the
cavity field operator and the $X^-$ dipole operator, respectively.
$\text{g}$  is the coupling strength between $X^-$ and the cavity mode.
$\kappa$,  $\kappa_s/2$, and $\gamma/2$ are  the cavity decay rate,
the side leakage (unwanted absorption) rate, and   the dipole decay
rate,  respectively. $\hat{a}_{in}$ and $\hat{a}_{in}'$ are the two
input field operators, and  $\hat{a}_r$ and $\hat{a}_t$ are the two
output field operators.  $\hat{H}$ and $\hat{G}$ are the noise
operators related to reservoirs. $\langle\sigma_z\rangle\approx-1$
is taken for a weak excitation approximation.

The complex reflection (transmission) coefficient given by
Eq.(\ref{eq4}) indicates that the reflected (transmitted) light
encounters a phase shift. When the QD is in the state
$|\uparrow\rangle$, the $|R^\uparrow\rangle$ or
$|L^\downarrow\rangle$  light feels the hot cavity ($\text{g}\neq0$)
and gets a phase shift $\varphi_h$  with the flip of the  photon
polarization and the photon propagation direction after reflection,
whereas the $|L^\uparrow\rangle$ or $|R^\downarrow\rangle$ light
feels the cold cavity ($\text{g}=0$) and gets a phase shift
$\varphi_0$  with the photon polarization and the photon propagation
direction unchanged after transmission. In the case that the QD is
in the state $|\downarrow\rangle$, the $|R^\uparrow\rangle$ or
$|L^\downarrow\rangle$  light feels the cold cavity  and gets a
phase shift $\varphi_0$  after transmission, whereas the
$|L^\uparrow\rangle$ or $|R^\downarrow\rangle$ light feels the hot
cavity and gets a phase shift $\varphi_h$   after reflection.  The
phase shift $\varphi_h$ or $\varphi_0$ can be adjusted by  the
frequency  detuning $\omega-\omega_0$
($\omega_c=\omega_{X^-}=\omega_0$).  When considering the
interaction with $\omega_{X^-}=\omega_c=\omega$, that is, the QD is
resonant with the cavity and  the spin of the independent electron
is connected by the resonant single photon,  the reflection and the
transmission coefficients for the uncoupled cavity (called a cold
cavity, $\text{g}$=0) and the coupled cavity (called a hot cavity,
$\text{g}\neq0$) can be simplified as
\begin{eqnarray}          \label{eq5}
r_0=\frac{\frac{\kappa_s}{2}}{\kappa+\frac{\kappa_s}{2}},\qquad
t_0=-\frac{\kappa}{\kappa+\frac{\kappa_s}{2}},\qquad
r=1+t,\qquad
t=-\frac{\frac{\gamma}{2}\kappa}{\frac{\gamma}{2}[\kappa+\frac{\kappa_s}{2}]+\text{g}^2}.
\end{eqnarray}
$r_0\rightarrow0$ and $t_0\rightarrow-1$ for the uncoupled cavity,
and $t\rightarrow 0$ and $r\rightarrow 1$ for the coupled cavity can
be achieved in the strong coupling regime $\text{g}>(\kappa,\gamma)$
in experiment by adjusting $|\omega-\omega_0|<\kappa$  when
$\kappa_s\ll\kappa$ (the ideal cavity)  and $\gamma=0.1\kappa$. That
is, if the photon couples to $X^-$, it will be reflected by the
cavity and both the propagation and the polarization of the photon
are flipped. If the photon does not couple to $X^-$, it will
transmit the cavity and acquire a $\pi$ mod $2\pi$ phase shift
relative to a reflected one \cite{Hu2,Appli1}. Therefore, the rules
of the input photon can be summarized as follows \cite{Appli1}:
\begin{eqnarray}         \label{eq1}
\begin{split}
&|R^\uparrow\uparrow\rangle\;\rightarrow\;|L^\downarrow\uparrow\rangle,\quad\;\;
|L^\uparrow\uparrow\rangle\;\rightarrow\;-|L^\uparrow\uparrow\rangle,\quad
|R^\downarrow\uparrow\rangle\;\rightarrow\;-|R^\downarrow\uparrow\rangle,\quad
|L^\downarrow\uparrow\rangle\;\rightarrow\;|R^\uparrow\uparrow\rangle,\\&
|R^\uparrow\downarrow\rangle\;\rightarrow\;-|R^\uparrow\downarrow\rangle,\;\;\;
|L^\uparrow\downarrow\rangle\;\rightarrow\;|R^\downarrow\downarrow\rangle,\quad\;\;
|R^\downarrow\downarrow\rangle\;\rightarrow\;|L^\uparrow\downarrow\rangle,\quad\;\;\;
|L^\downarrow\downarrow\rangle\;\rightarrow\;-|L^\downarrow\downarrow\rangle.
\end{split}
\end{eqnarray}
The photon spin $s_z$ remains unchanged upon the reflection as the circular polarization of the photons is dependent of their propagation direction, so $|R^\uparrow\rangle$ and $|L^\downarrow\rangle$ with the same $s_z=+1$, $|L^\uparrow\rangle$ and $|R^\downarrow\rangle$ with $s_z=-1$.

\begin{figure}[!h]
\centering\includegraphics[width=9 cm,angle=0]{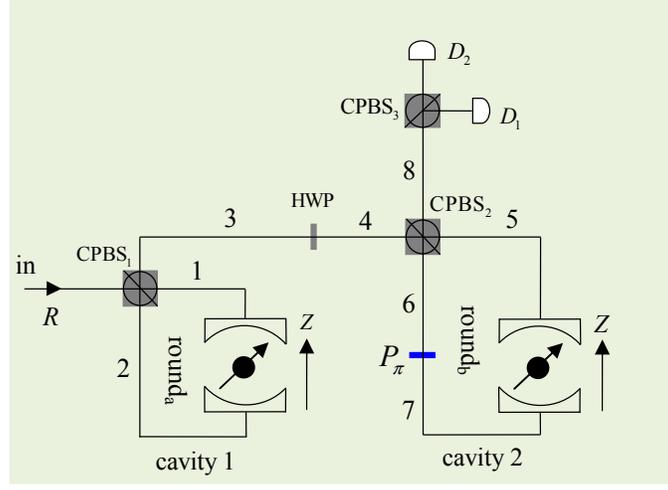}
\caption{Schematic diagram for compactly implementing a CNOT gate on
two electron-spin qubits in optical microcavities. The electron spin
in cavity 1 is the control qubit and that in cavity 2 is the target
qubit.  CPBS$_i$ ($i=1,2$) is a polarizing beam splitter (PBS) in
the circular basis $\{\vert R\rangle, \vert L\rangle\}$, which
transmits the right-circularly polarized photon $|R\rangle$ and
reflects the left-circularly polarized photon $|L\rangle$,
respectively. CPBS can be constructed by a PBS in the horizontal and
vertical basis followed by a quarter-wave plate or a half-wave plate
(HWP) whose optical axis is set at 22.5$^\circ$. Here HWP is used to
implement a Hadamard ($H^p$) operation on the polarization photon
passing through it. $P_\pi$ is a phase shifter which contributes a
$\pi$ phase shift on the photon passing through it. D$_1$ and D$_2$
are two single-photon detectors. \label{figure2}}
\end{figure}

\bigskip

\textbf{Compactly implementing CNOT gate on a two-qubit
electron-spin system.} Now, let us describe the construction of a deterministic CNOT gate
on the two stationary electron-spin qubits assisted by double-sided
QD-cavity systems. It flips the target qubit if and only if (iff) the control qubit is in $|\downarrow\rangle$. That is,
\begin{eqnarray}
\text{CNOT}=\left(\begin{array}{ccc}
I_2&0&0\\
0&0&1\\
0&1&0\\
\end{array}\right)
\end{eqnarray}
in the basis $\{|\uparrow\uparrow\rangle,|\uparrow\downarrow\rangle,|\downarrow\uparrow\rangle,|\downarrow\downarrow\rangle\}$. Here $I_2$ is a $2\times2$ identity matrix. Suppose the two remote electrons confined in
cavities 1 and 2 are  initially prepared in  the  state
\begin{eqnarray}                    \label{eq6}
|\psi_{\rm{in}}\rangle_{ct}=
\frac{1}{\sqrt{2}}(|\uparrow\rangle_{c}+|\downarrow\rangle_{c})\otimes
(\cos\alpha|\uparrow\rangle_{t}+\sin\alpha|\downarrow\rangle_{t}).
\end{eqnarray}
The subscript $c$  represents the control qubit confined in  cavity 1
and $t$ stands for the target qubit confined in cavity 2. 


Our schematic diagram  for a CNOT gate on the two stationary
electron-spin qubits is shown in Fig. \ref{figure2}. The input
single photon in the state $|R^\downarrow\rangle$ transmits the
polarizing beam splitter in the circular basis CPBS$_{1}$ to the
spatial mode 1 and injects into  cavity 1 which induces the
transformations
$|R^\downarrow\rangle_1|\uparrow\rangle_c\xrightarrow{\text{cavity}}-|R^\downarrow\rangle_2|\uparrow\rangle_c$
and
$|R^\downarrow\rangle_1|\downarrow\rangle_c\xrightarrow{\text{cavity}}|L^\uparrow\rangle_1|\downarrow\rangle_c$.
The subscript $i$  ($i=1,2,\cdots,$) of $|R\rangle$ ($|L\rangle$)
represents the spatial mode $i$ from which the photon is emitted.
CPBS can be constructed by a PBS in the horizontal and vertical
basis followed by a half-wave plate (HWP)  at 22.5$^\circ$.
Subsequently, the photon in the state $|R^\downarrow\rangle_2$ or
$|L^\uparrow\rangle_1$ is led to the spatial mode 3 by CPBS$_1$
which completes the transformations
$|R^\downarrow\rangle_2\xrightarrow{\text{CPBS}_1}|R^\downarrow\rangle_3$
and
$|L^\uparrow\rangle_1\xrightarrow{\text{CPBS}_1}|L^\uparrow\rangle_3$.
That is,  round$_a$ transforms the state of the whole system
composed of the two excess electrons inside  cavities 1 and 2 and
the single photon from the initial state $|\psi_{0}\rangle$ to
$|\psi_{1}\rangle$. Here
\begin{eqnarray}                      \label{eq7}
|\psi_{0}\rangle=|R^\downarrow\rangle\otimes|\psi_{\rm{in}}\rangle_{ct},
\end{eqnarray}
\begin{eqnarray}                      \label{eq8}
|\psi_1\rangle=\frac{1}{\sqrt{2}}(-|R^\downarrow\rangle_3|\uparrow\rangle_c+|L^\uparrow\rangle_3|\downarrow\rangle_c)(\cos\alpha|\uparrow\rangle_t+\sin\alpha|\downarrow\rangle_t).
\end{eqnarray}
The transformations of round$_a$ can be described by the following
unitary matrix:
\begin{eqnarray}                      \label{eq9}
U_{\text{round}_a}=\left(\begin{array}{cccc}
-1&0&0&0\\
0&0&0&1\\
0&0&-1&0\\
0&1&0&0\\
\end{array}\right)
\end{eqnarray}
in the basis $\{$$|R^\downarrow\rangle|\uparrow\rangle$,
$|R^\downarrow\rangle|\downarrow\rangle$,
$|L^\uparrow\rangle|\uparrow\rangle$, $|L^\uparrow\rangle|\downarrow
\rangle$$\}$. Before the photon in the state
$|R^{\downarrow}\rangle_{3}$ or $|L^{\uparrow}\rangle_{3}$ arrives
at CPBS$_{2}$ simultaneously,  a Hadamard operation $H^p$ is
performed on it (i.e., let it pass  through the HWP whose optical
axis is set at 22.5$^\circ$) and an $H^{e}$ operation is
performed on the excess electron inside  cavity 2 before and after the
photon interacts with the QD in cavity 2. Here an $H^{p}$ operation
completes the transformations
\begin{eqnarray}                     \label{eq10}
|R\rangle\xrightarrow{H^p}\frac{1}{\sqrt2}(|R\rangle+|L\rangle),\qquad\quad
|L\rangle\xrightarrow{H^p}\frac{1}{\sqrt2}(|R\rangle-|L\rangle),
\end{eqnarray}
and an $H^{e}$ operation completes the transformations
\begin{eqnarray}                     \label{eq11}
|\uparrow\rangle\xrightarrow{H^e}\frac{1}{\sqrt2}(|\uparrow\rangle+|\downarrow\rangle),\qquad\quad
|\downarrow\rangle\xrightarrow{H^e}\frac{1}{\sqrt2}(|\uparrow\rangle-|\downarrow\rangle).
\end{eqnarray}
The operations ($H^{p},\;H^{e}\rightarrow\text{round}_b\rightarrow
H^e$) induce the state of the whole system  to be
\begin{eqnarray}                   \label{eq12}
\begin{split}
|\psi_2\rangle =&
-\frac{1}{2}(|R^\downarrow\rangle_8+|L^\uparrow\rangle_8)|\uparrow\rangle_c(\cos\alpha|\uparrow\rangle_t+\sin\alpha|\downarrow\rangle_t)\\&
+\frac{1}{2}(|R^\downarrow\rangle_8-|L^\uparrow\rangle_8)|\downarrow\rangle_c(\cos\alpha|\downarrow\rangle_t+\sin\alpha|\uparrow\rangle_t).
\end{split}
\end{eqnarray}
Here round$_b$  completes the transformation
\begin{eqnarray}       \label{eq13}
\begin{split}
&|R^\downarrow\rangle_4|\uparrow\rangle\;\xrightarrow{\rm{round}_b}\;|R^\downarrow\rangle_8|\uparrow\rangle,\;\;\;\;
|R^\downarrow\rangle_4|\downarrow\rangle \;\xrightarrow{\rm{round}_b}\;|L^\uparrow\rangle_8|\downarrow\rangle,\\
&|L^\uparrow\rangle_4|\uparrow\rangle\;\xrightarrow{\rm{round}_b}\;|L^\uparrow\rangle_8|\uparrow\rangle,\;\;\;\;\,
|L^\uparrow\rangle_4|\downarrow\rangle\;\xrightarrow{\rm{round}_b}\;|R^\downarrow\rangle_8|\downarrow\rangle.
\end{split}
\end{eqnarray}

From Eq.(\ref{eq12}), one can see that if the output photon is in
the state $|L\rangle$, the CNOT gate is accomplished. If the output
photon is in the state $|R\rangle$, after a   feed-forward
single-qubit operation
$-\sigma_z=-|\uparrow\rangle\langle\uparrow|+|\downarrow\rangle\langle\downarrow|$
is performed on the QD in cavity 1, the CNOT gate is accomplished as
well. That is, after the measurement on the output photon and the
feed-forward operation on the QD, the state of the two-electron
system becomes
\begin{eqnarray}                  \label{eq14}
|\psi_{\rm{out}}\rangle_{ct}=
  \frac{|\uparrow\rangle_c}{\sqrt{2}}(\cos\alpha|\uparrow\rangle_t + \sin\alpha|\downarrow\rangle_t)
  +\frac{|\downarrow\rangle_c}{\sqrt{2}}(\cos\alpha|\downarrow\rangle_t +  \sin\alpha |\uparrow\rangle_t).
\end{eqnarray}
Therefore, the quantum circuit shown in Fig. \ref{figure2} converts
the input state $|\psi_{\rm{in}}\rangle_{ct}$ to the output state
$|\psi_{\rm{out}}\rangle_{ct}$, i.e.,
$|\psi_{\rm{in}}\rangle_{ct}\xrightarrow{\text{CNOT}}|\psi_{\rm{out}}\rangle_{ct}$.
It implements a deterministic electron-spin CNOT gate which flips
the state of the target electron-spin qubit when the control
electron-spin qubit is in the state $|\downarrow\rangle$; otherwise,
nothing is done on the target qubit.
\\

\begin{figure}[!h]
\centering
\includegraphics[width=8 cm,angle=0]{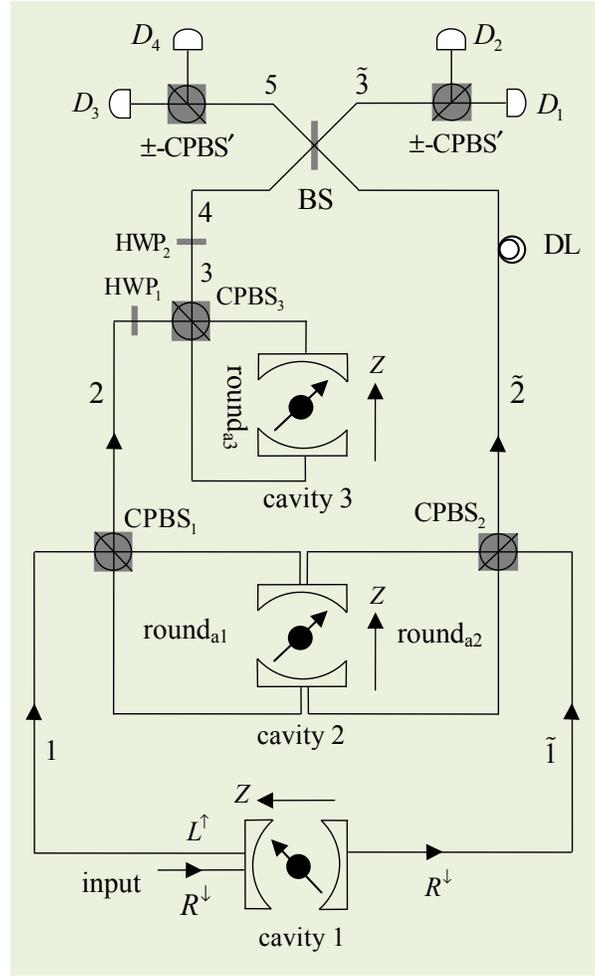}
\caption{Schematic diagram  for compactly implementing a
deterministic Toffoli gate on three electron-spin qubits, assisted
by  QD-cavity systems. The electron spins in cavities 1 and 2 are
the two control qubits $c_1$ and $c_2$, and the spin in cavity 3 is
the target qubit $t$. DL is a time-delay device which makes the
photon emitting from the spatial mode 4 interfere with the photon
emitting from the spatial mode $\tilde{2}$ at the 50:50 beam splitter
(BS). The polarizing beam
splitter  in the $\{\pm\}$ circular basis ($\pm$--PBS') transmits the
diagonal-polarization photon
$|+\rangle=(|R\rangle+|L\rangle)/{\sqrt{2}}$ and reflects the
antidiagonal-polarization photon
$|-\rangle=(|R\rangle-|L\rangle)/{\sqrt{2}}$, respectively.
\label{figure3}}
\end{figure}

\bigskip

\textbf{Toffoli gate on a three-qubit electron-spin  system.}  The
schematic diagram of our scheme for compactly implementing a
three-qubit electron-spin Toffoli gate is shown in Fig.
\ref{figure3}. It implements a NOT operation on the target qubit iff
both the two control qubits are in $|\downarrow\rangle$.  That is,
the unitary transformation of the Toffoli gate on the three QDs can
be characterized by the following matrix
 \begin{eqnarray}
\text{Toffoli}=\left(\begin{array}{ccc}
I_6&0&0\\
0&0&1\\
0&1&0\\
\end{array}\right)
\end{eqnarray}
in the basis $\{|\uparrow\uparrow\uparrow\rangle,|\uparrow\uparrow\downarrow\rangle,|\uparrow\downarrow\uparrow\rangle,
|\uparrow\downarrow\downarrow\rangle, |\downarrow\uparrow\uparrow\rangle,|\downarrow\uparrow\downarrow\rangle,|\downarrow\downarrow\uparrow\rangle,
|\downarrow\downarrow\downarrow\rangle\}$.  Here $I_6$ is a $6\times6$ identity matrix.
Suppose the spins of the three excess electrons in  cavities 1, 2, and 3 are
encoded as the first control, the second control, and the target
qubits, respectively. The system composed of those three electrons
is initially prepared in the state
\begin{eqnarray}                      \label{eq15}
|\Xi_{\rm{in}}\rangle_{c_1
c_2t}=\frac{1}{\sqrt{2}}(|\uparrow\rangle_{c_1}+|\downarrow\rangle_{c_1})\otimes\frac{1}{\sqrt{2}}(|\uparrow\rangle_{c_2}+|\downarrow\rangle_{c_2})\otimes(\cos\alpha|\uparrow\rangle_{t}+\sin\alpha|\downarrow\rangle_{t}).
\end{eqnarray}
Our protocol for a Toffoli gate can be achieved with  three steps
discussed in detail as follows.

First, a single photon in the state $|R^\downarrow\rangle$ is
injected into the input port, shown in Fig. \ref{figure3}. If the
photon is transmitted through cavity 1, it is emitted from the
spatial mode $\tilde{1}$ with the state
$-|R^\downarrow\rangle_{\tilde{1}}$. If the photon is reflected by
cavity 1, it is emitted from  the spatial mode 1 with the state
$|L^\uparrow\rangle_1$. The nonlinear interaction between the input
photon and the QD in cavity 1 transforms the state of the composite
system composed of the three electrons ($c_1$, $c_2$ and $t$) and
the single photon into
\begin{eqnarray}                      \label{eq16}
|\Xi_1\rangle=
\frac{1}{2}(-|R^\downarrow\rangle_{\tilde{1}}|\uparrow\rangle_{c_1}+|L^\uparrow\rangle_{1}|\downarrow\rangle_{c_1})(|\uparrow\rangle_{c_2}+|\downarrow\rangle_{c_2})(\cos\alpha|\uparrow\rangle_t+\sin\alpha|\downarrow\rangle_t).
\end{eqnarray}
When the photon is emitted from the spatial mode 1, it will be
injected into round$_{a1}$ described by Eq.(\ref{eq9}). When the
photon is emitted from the spatial mode $\tilde{1}$, it will be
injected into round$_{a2}$ described by Eq.(\ref{eq9}).
Round$_{a1}$ and round$_{a2}$ transform the state of the whole
system into
\begin{eqnarray}                      \label{eq17}
\begin{split}
|\Xi_2\rangle=&
\frac{|\uparrow\rangle_{c_1}}{2}(|R^\downarrow\rangle_{\tilde{2}}|\uparrow\rangle_{c_2}-|L^\uparrow\rangle_{\tilde{2}}|\downarrow\rangle_{c_2})(\cos\alpha|\uparrow\rangle_t+\sin\alpha|\downarrow\rangle_t)\\
&
-\frac{|\downarrow\rangle_{c_1}}{2}(|L^\uparrow\rangle_{2}|\uparrow\rangle_{c_2}-|R^\downarrow\rangle_{2}|\downarrow\rangle_{c_2})(\cos\alpha|\uparrow\rangle_t+\sin\alpha|\downarrow\rangle_t).
\end{split}
\end{eqnarray}
The photon emitting from the spatial mode $\tilde{2}$ does not
interact with the QD in cavity 3, while the photon  emitting
from the spatial mode 2 is injected into  cavity 3. Before and after
the photon emitting from  the spatial mode 2 interacts with the QD
in cavity 3, an $H^p$ is performed on it with HWP$_1$
and HWP$_2$, and
 an $H^e$  is performed on the electron
in cavity 3, respectively. Operations
($\text{HWP}_{1},H^e\rightarrow \text{round}_{a3}\rightarrow
\text{HWP}_{2},H^e$) transform the state of the whole system into
\begin{eqnarray}                      \label{eq18}
\begin{split}
|\Xi_3\rangle=&
\frac{1}{2}(|R^\downarrow\rangle_{\tilde{2}}|\uparrow\rangle_{c_1}|\uparrow\rangle_{c_2}
-|L^\uparrow\rangle_{\tilde{2}}|\uparrow\rangle_{c_1}|\downarrow\rangle_{c_2}
+|L^\uparrow\rangle_{4}|\downarrow\rangle_{c_1}|\uparrow\rangle_{c_2})(\cos\alpha|\uparrow\rangle_t+\sin\alpha|\downarrow\rangle_t)\\&
-\frac{1}{2}|R^\downarrow\rangle_{4}|\downarrow\rangle_{c_1}|\downarrow\rangle_{c_2}(\cos\alpha|\downarrow\rangle_t+\sin\alpha|\uparrow\rangle_t).
\end{split}
\end{eqnarray}

Second, the photon emitting from the spatial mode 4 interferes with
the photon emitting from the spatial mode $\tilde{2}$ at the 50:50
beam splitter (BS) which completes the transformations
\begin{eqnarray}                     \label{eq19}
\begin{split}
&|R\rangle_4\xrightarrow{\rm{BS}}\frac{1}{\sqrt2}(|R\rangle_5+|R\rangle_{\tilde{3}}),\qquad
|L\rangle_4 \xrightarrow{\rm{BS}} \frac{1}{\sqrt2}(|L\rangle_5+|L\rangle_{\tilde{3}}),\\
&|R\rangle_{\tilde{2}}\xrightarrow{\rm{BS}}\frac{1}{\sqrt2}(|R\rangle_5-|R\rangle_{\tilde{3}}),\qquad
|L\rangle_{\tilde{2}}\xrightarrow{\rm{BS}}\frac{1}{\sqrt2}(|L\rangle_5-|L\rangle_{\tilde{3}}).
\end{split}
\end{eqnarray}

Third, the output photon is measured in the basis
$\{|\pm\rangle=(|R\rangle\pm|L\rangle)/\sqrt{2}\}$. After some
feed-forward single-qubit operations are performed on the electron-spin qubits,  a
Toffoli gate on the three-qubit electron-spin system is
achieved. That is, the state of the system composed of the three
electrons confined in QDs becomes
\begin{eqnarray}                      \label{eq20}
\begin{split}
|\Xi_{\rm{out}}\rangle_{c_1c_2t}=&
\frac{1}{2}(|\uparrow\rangle_{c_1}|\uparrow\rangle_{c_2}+|\uparrow\rangle_{c_1}|\downarrow\rangle_{c_2}+|\downarrow\rangle_{c_1}|\uparrow\rangle_{c_2})(\cos\alpha|\uparrow\rangle_t+\sin\alpha|\downarrow\rangle_t)\\&
+\frac{1}{2}|\downarrow\rangle_{c_1}|\downarrow\rangle_{c_2}(\cos\alpha|\downarrow\rangle_t+\sin\alpha|\uparrow\rangle_t).
\end{split}
\end{eqnarray}
Here, the response of detector $D_2$ indicates that the Toffoli gate
is successful. The response of detector $D_1$ indicates that the
feed-forward single-qubit operations $-\sigma_z$ and $\sigma_z$
should be performed on the QDs in  cavities 1 and 2, respectively.
The response of detector $D_3$ ($D_4$) indicates that $\sigma_z$
should be performed on the QD in  cavity 2 (1).

From Eq.(\ref{eq20}), one can see that the quantum circuit shown in
Fig. 3 converts the input state $|\Xi_{\rm{in}}\rangle_{c_1 c_2 t}$
to the output state $|\Xi_{\rm{out}}\rangle_{c_1 c_2 t}$, i.e.,
$|\Xi_{\rm{in}}\rangle_{c_1 c_2 t}
\xrightarrow{\rm{Toffoli}}|\Xi_{\rm{out}}\rangle_{c_1 c_2 t}$. That
is, it can be used to implement a Toffoli gate on a three-qubit
electron-spin system, which flips the state of the target
electron-spin qubit iff both the two control electron-spin qubits
are in the state $|\downarrow\rangle$, in a deterministic way.


\bigskip

{\large \textbf{Discussion}}

We have discussed the construction of universal quantum gates on
electron-spin qubits, assisted by double-sided optical
microcavities. In above discussion, all the QD-cavity systems are
ideal. That is, the side leakage of the cavities is not taken into
account. However, there inevitably  exists the side leakage (which
includes the material background absorption and the cavity loss) in
a realistic QD-cavity system, which induces the polarize-bit-flip
errors and different amplitudes between the coupled and the
uncoupled photons. If the cavity leak is taken into account, the
optical selection rules employed in our work become
\cite{Hu2,Appli1}
\begin{eqnarray} \label{eq21}
\begin{split}
&|R^\downarrow\downarrow\rangle  \rightarrow
|r||L^\uparrow\downarrow\rangle+|t||R^\downarrow\downarrow\rangle,\qquad\quad\;\;
|L^\uparrow  \downarrow\rangle  \rightarrow   |r||R^\downarrow\downarrow\rangle+|t||L^\uparrow \downarrow\rangle,\\
&|R^\downarrow\uparrow\rangle    \rightarrow
-|t_0||R^\downarrow\uparrow\rangle-|r_0||L^\uparrow\uparrow\rangle,\qquad
|L^\uparrow  \uparrow\rangle    \rightarrow   -|t_0||L^\uparrow
\uparrow\rangle-|r_0||R^\downarrow \uparrow\rangle.
\end{split}
\end{eqnarray}

QDs inside microcavities with   high quality factors $Q$ are of
particular interest for studying light-matter interaction. The
photon loss strongly reduces $Q$. In micropillar microcavities, a
drop of $Q$ takes place  with the pillar diameter $d$ due to an
increasing photon loss \cite{observe2}. It is desired to increase
the $Q$ values but maintain a small effective optical mode volume,
which can be achieved by improving the sample design, growth, and
structure \cite{observe2}. Some coupling strengths and the quality
factors of the QD-cavity system have been experimentally achieved in
various microcavities and nanocavities
\cite{observe2,observe1,observe3,observe4,Largesize} (see Tab. 1).

\begin{table}
\centering
\caption{The quality factors and coupled strengths of the QD-cavity systems have been achieved in experiments.}
\begin{tabular}{ccc}
\hline  \hline

coupled strength                           &$\;$ quality factor      &  $\;$ parameters     \\
\hline
$g/(\kappa+\kappa_s)\simeq0.5$            &  $Q\sim 8800$            & $d\;\sim 1.5\; \mu m$  \cite{observe1}   \\
$g/(\kappa+\kappa_s)\simeq2.4$             & $Q\sim 4\times 10^4$    &    $d\sim1.5 \;\mu m$  \cite{observe2} \\
$g/(\kappa+\kappa_s)\simeq 0.8$  &  $Q\sim6.5\times10^4$             &  $d\sim 7.3\; \mu m$ \cite{Largesize}                 \\
$g/(\kappa+\kappa_s)\simeq 1.0$    &   $Q\simeq 1.7 \times 10^4$     & $d\;\sim 1.5$,\; $\kappa/\kappa_s\simeq 0.7$  \cite{Hu4}  \\

                             \hline  \hline
\end{tabular}\label{table}
\end{table}

Quantum logic gates play an important role in quantum computing. The
feasibilities of realizing universal quantum computation with
superconducting qubits in circuit-quantum-electrodynamics setups
have been investigated \cite{superPRL,superPRB}. Romero \emph{et
al.} \cite{superPRL} proposed a scheme for realizing an ultrafast
controlled-phase gate in current circuit-QED technology at
subnanosecond time scales with the fidelity of the gate $F=99\%$.
Stojanovi\'{c} \emph{et al.} \cite{superPRB} designed a quantum
circuit for directly and fast realizing a Toffoli gate on
superconducting qubits within 75 ns with $F>90\%$, and within 140 ns
with $F>99\%$. Based on specific solid-state platforms, proposals
for realizing the CNOT and $\sqrt{\text{SWAP}}$ (SWAP) gates in
two-qubit Heisenberg spin chains have been proposed
\cite{Heisen1,Heisen3,Heisen4}.

In our work, the schemes for quantum gates based on the electron
spins in QDs are particularly interesting because of their good
scalability and long coherence time which can be extended from
$T_2\sim $ ns range \cite{cohertime1,cohertime2,cohertime3} to
$T_2\sim$ $\mu$s range using the spin echo technique. The
weak-excitation approximation is taken in QD-cavity systems, and it
demands the number of the intracavity photons less than the number
of the critical photons \cite{critical} $n_0=\gamma^2/2g^2$.  That
is to say, the time interval between two intracavity photons should
be longer than $\tau/n_0\sim$ ns (By taking
$g/(\kappa+\kappa_s)=1.0$, $\kappa_s/\kappa=0.7$ and
$\gamma=0.1\kappa$ for a micropillar microcavity with diameter
$d=1.5 \mu$ m, $Q=1.7\times10^4$, one can get $n_0=2\times10^{-3}$,
$\tau=9$ ps, and  $\tau/n_0=4.5 $ ns).  Here $\tau$ is the cavity
photon lifetime and it is around 10 ps. In our schemes, we need only
one single photon. The speed of the photon interacting with the
electron-spin is determined by the cavity photon lifetime. Moreover,
the photon medium is easy to be controlled and manipulated.

In summary, we have proposed two deterministic schemes for compactly
implementing a set of universal quantum gates on stationary
electron-spin qubits, including the two-qubit CNOT gate and
three-qubit Toffoli gate. Our universal quantum gates are scalable.
Different to the hybrid schemes \cite{Hu1,Hu2,Hu3,Hu4,Appli1} acting
on the photon-electron qubits, our schemes are based on solid-state
systems (QD-cavity systems).
 Moreover, our schemes do
not require additional electron-spin qubits. Comparing with the
synthesis of gates in terms of CNOT gate, our scheme for Toffoli
gate is powerful. It is required 6 CNOT gates to synthesize a
Toffoli gate in the best case \cite{Toffolicost}. It is worth
pointing out that with the present technology, our schemes are
feasible. Both high fidelities and high efficiencies can be achieved
when the ratio of the side leakage to the cavity decay is low.

With universal quantum gates on electron-spin qubits, scalable
quantum computing can be achieved. Maybe it is interesting to
investigate some important quantum algorithms based on electron-spin
systems in future.

\bigskip

{\large \textbf{Methods}}

\textbf{Manipulation and measurement of the QD spin.} The QD-spin
superposition state can be prepared by performing single spin-qubit
rotations with picosecond optical pulses
\cite{spin-manipulate1,spin-manipulate2} or nanosecond electron-spin
resonance (ESR) microwave pulses in an external magnetic field
\cite{cohertime1,cohertime2} on the spin eigenstate which is
prepared by an optical pumping or optical cooling
\cite{superposition1,superposition2}. Ultrafast optical coherent
manipulation of a QD-spin qubit has been demonstrated in a
picosecond or femtosecond time scale
\cite{spin-manipulate1,spin-manipulate3}, and an ultrafast $\pi/2$
spin rotation can be used to complete a Hadamard operation on a spin
qubit. In our schemes, the spin level levels of $X^-$ are in
degeneracy as the anisotropic electron-hole exchange interaction,
which lifts the degeneracy of the neutral exciton
\cite{fine-structure1,fine-structure2}, vanishes  for  the charged
excitons in self-assembled QDs. The ESR-based (Faraday geometry) and
coherent optical (geometric phase or AC stark shift) QD spin
manipulation require an application of a magnetic field which lifts
the  sublevels of the QD spin, and the degeneracy transition rules
given by Eq.(\ref{eq1}) are not valid any more as the $R$- and
$L$-polarized transitions have a small energy difference. In our
schemes, the single-qubit operations are performed on the QDs before
or after the single photon interacts with the QDs, so the single
photon can interact with the QDs in the absence of the external
magnetic field. This trick might be employed to overcome above
drawback. Experimentally demonstrating this trick is a challenge
with current experimental technology as the required
magnetic-relaxation timescales $\sim$ ms is a challenge, and the
accurate control of the switch on the required timing needs to be
considered. Furthermore, the electron coherence time without a
magnetic field is much shorter than that in the presence of a
magnetic field (another possible solution to achieve the degeneracy
transitions in an external magnetic field is to employ the QDs with
the identical $\text{g}$ factors of the electron and hole). From the
rules given by Eq.(\ref{eq1}), one can see that a single photon can
be used to complete a 180$^\circ$ spin rotation of the QD around the
optical axis and detect the polarization of the QD spin. Take the
$|R^\downarrow\rangle$ light as an example, the nonlinear
interaction between such a single photon and the QD induces the
transformation
\begin{eqnarray} \label{eq22}
|R^\downarrow\rangle(\alpha|\uparrow\rangle+\beta|\downarrow\rangle)\rightarrow-\alpha|R^\downarrow\rangle|\uparrow\rangle+\beta|L^\uparrow\rangle|\downarrow\rangle.
\end{eqnarray}
The single photon in the state $|R\rangle$ ($|L\rangle$) indicates that the QD is in the state $|\uparrow\rangle$ ($|\downarrow\rangle$).

\bigskip

\textbf{The average fidelities of the gates.} In our work, the fidelity of a universal quantum gate is defined as
$F=|\langle\Psi_r|\Psi_i\rangle|^2$. Here $\vert \Psi_r\rangle$
presents the finial state in a realistic QD-cavity system composed
of excess electrons encoded for the gates and a single photon
medium, whereas $\vert \Psi_i\rangle$ represents the final state of
this composite system in the ideal condition. The reflection and
transmission coefficients of the QD-cavity system for a realistic
system given by Eqs.(\ref{eq5}) and (\ref{eq6}) affect the
fidelities of our universal quantum gates. Taking the CNOT gate as
an example, the average fidelity of the CNOT gate is given by
\begin{eqnarray}                  \label{eq23}
\overline{F}_{CT}=\frac{1}{2\pi}\int_0^{2\pi}d\alpha|\langle\psi_r|\psi_i\rangle|^2.
\end{eqnarray}
From above discussion, one can see that the output state of our
scheme for the CNOT gate in the ideal case $|\psi_{\rm{i}}\rangle$
can be expressed as
\begin{eqnarray}                  \label{eq24}
\begin{split}
|\psi_{i}\rangle=
 \frac{|-\rangle}{\sqrt{2}}(\cos\alpha|\downarrow\rangle_c|\downarrow\rangle_t +\sin\alpha |\downarrow\rangle_c|\uparrow\rangle_t)-
  \frac{|+\rangle}{\sqrt{2}}(\cos\alpha|\uparrow\rangle_c|\uparrow\rangle_t + \sin\alpha|\uparrow\rangle_c|\downarrow\rangle_t).
\end{split}
\end{eqnarray}
If we consider each input-output process of our schemes in the
real case described by  Eq.(\ref{eq21}), the output state of our
scheme for the CNOT gate can be rewritten as
\begin{eqnarray}                  \label{eq25}
\begin{split}
|\psi_{r}\rangle=&
\frac{|+\rangle}{2\sqrt{2}}\big[|t|(\xi\cos\alpha+\zeta\sin\alpha)|\downarrow\rangle_c|\uparrow\rangle_t+
|t|(\zeta\cos\alpha+\xi\sin\alpha)|\downarrow\rangle_c|\downarrow\rangle_t\\&
-|t_0|(\xi\cos\alpha+\zeta\sin\alpha)|\uparrow\rangle_c|\uparrow\rangle_t
-|t_0|(\zeta\cos\alpha+\xi\sin\alpha)|\uparrow\rangle_c|\downarrow\rangle_t\big]\\&+
\frac{|-\rangle}{\sqrt{2}}\big[|r||\downarrow\rangle_c(\cos\alpha|\downarrow\rangle_t+\sin\alpha|\uparrow\rangle_t)-|r_0||\uparrow\rangle_c(\cos\alpha|\downarrow\rangle_t+\sin\alpha|\uparrow\rangle_t)\big]
\end{split}
\end{eqnarray}
with $\xi=(|t_0|-|r_0|)-(|t|-|r|)$ and
$\zeta=(|t_0|-|r_0|)+(|t|-|r|)$.

\bigskip

\textbf{The efficiencies of the gates.}  The efficiency of a quantum
gate is defined as the yield of the photons
($\eta=n_{\rm{output}}/n_{\rm{input}}$), that is, the ratio of the
number of the output photons $n_{\rm{output}}$ to that of the input photons $n_{\rm{input}}$. It is
also sensitive to the reflection and transmission coefficients of
the QD-cavity system. The efficiencies of the CNOT gate and Toffoli gate
can be respectively expressed as
\begin{eqnarray}                \label{eq26}
\eta_{CT}=(1-|t_0||r_0|-|t||r|)^2,
\end{eqnarray}
\begin{eqnarray}             \label{eq27}
\eta_{T}=\frac{1}{2}(1-|t_0||r_0|-|t||r|)^2(2-|t_0||r_0|-|t||r|)).
\end{eqnarray}

\begin{figure}[htb]                    
\centering
\includegraphics[width=6.2 cm]{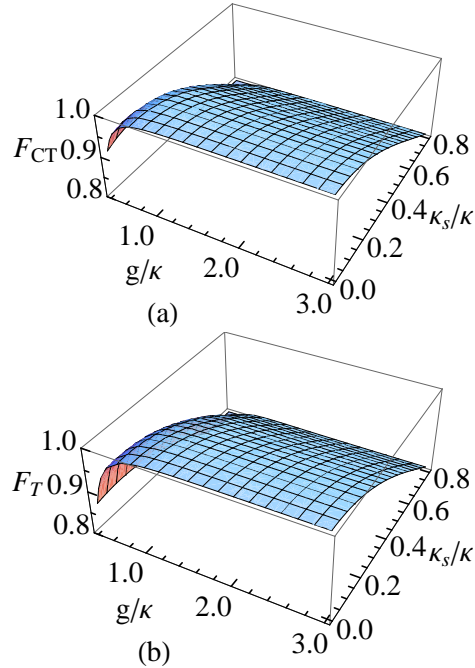}
\caption{The fidelities of the present  deterministic scalable
universal quantum gates on electron-spin systems vs the coupling
strength $g/\kappa$ and the side leakage rate $\kappa_s/\kappa$. (a)
The fidelity of the CNOT gate $F_{CT}$; (b) the fidelity of the
Toffoli gate $F_{T}$. $\omega_c=\omega_{x^-}=\omega$ and
$\gamma=0.1\kappa$ are taken for (a) and (b) as $\gamma$ is about
several $\mu eV$ in experiment. \label{figure4}}
\end{figure}

\begin{figure}[htb]                    
\centering
\includegraphics[width=6.2 cm]{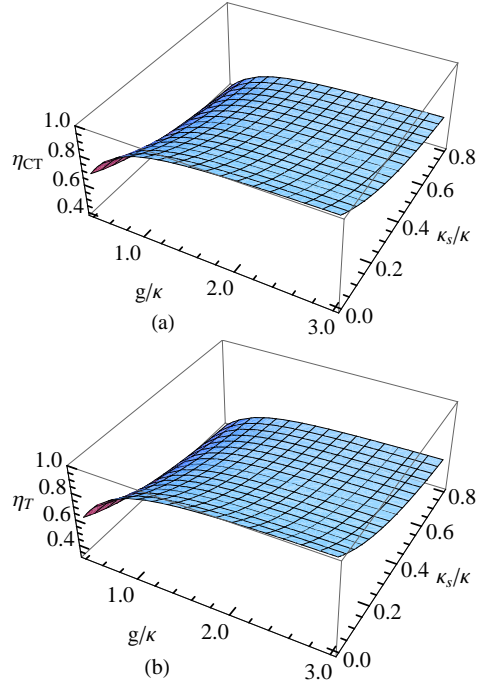}
\caption{The efficiencies of the present universal quantum gates vs
the coupling strength $g/\kappa$ and the side leakage rate
$\kappa_s/\kappa$. (a) The efficiency of the CNOT gate $\eta_{CT}$;
(b) the efficiency of the Toffoli gate $\eta_{T}$.
$\omega_c=\omega_{x^-}=\omega$ and $\gamma=0.1\kappa$ are taken for
(a) and (b).\label{figure5}}
\end{figure}

\bigskip

\textbf{Fidelity and efficiency  estimation.} The average fidelities
$F$ and the efficiencies $\eta$ of the present universal quantum
gates as the function of the coupling strength $g/\kappa$ and the
side leakage rate $\kappa_s/\kappa$ are shown in Figs. \ref{figure4}
and \ref{figure5}, respectively. These results indicate that the
fidelity and the efficiency behaviors of the two gates are similar
to each other. When $\kappa_s/\kappa$ is very small and $g/\kappa$
is large, the fidelities and the efficiencies of the gates are close
to one. Both the fidelities and the efficiencies are high in both
the strong coupling regime [$g>(\kappa,\gamma)$] and the weak
coupling regime [$g<(\kappa,\gamma)$] when the ratio of the side
leakage to the cavity decay is low.  However, with an increase
$\kappa_s/\kappa$ or a decrease $g/\kappa$,  the fidelities and the
efficiencies of the gates declines. Here the fidelities and the
efficiencies of the single-qubit operations performed on the photons
or the QDs in our schemes are  unity, that is, the imperfect
operation and the photon loss in the single-qubit operations are not
taken into account. To get high fidelities and efficiencies of the
gates, a small side leakage is required, and it can be achieved by
improving the sample growth and optimizing etch processing
\cite{observe2}. $\kappa_s/\kappa=0.05$ may be achieved for a $Q
\sim 165000$ pillar microcavity \cite{observe2} and reduce $Q $ to
$\sim 9000$ by decreasing  the reflection of the top mirror
\cite{Hu1}. For a QD-cavity system,
 $g/(\kappa+\kappa_s)\simeq0.5$, $g/\kappa=2.4$, and $g/(\kappa+\kappa_s)\simeq 0.8$ have been observed in experiment \cite{observe1,observe2,Largesize}. For our
protocols, when $g/\kappa=2.4$ and $\kappa_s/\kappa=0.2$,
$F_{CT}=0.981517$ and  $F_{T}=0.983317$ with $\eta_{CT}=0.82597$ and
$\eta_{T}=0.788322$, respectively. If the cavity side leakage is negligible, the average
fidelities of our quantum gates are close to one with a near-unity success probability ($F_{CT}=0.999916$, $F_{T}=0.999857$,
$\eta_{CT}=0.98301$, $\eta_{T}=0.978816$ when $\kappa_s=0$). Here
the subscripts $CT$ and $T$ represent our CNOT gate and Toffoli
gate, respectively.

The fidelities of aforementioned universal quantum gates are
decreased by a amount of $ 1-\exp(-\tau/T_2)$ due to the exciton
dephasing effect caused by the exciton decoherence \cite{Hu2}. That
is, the fidelity depends on  the $X^-$ coherence time  $T_2$ and the
cavity photon coherence time $\tau$. Since the information of the
polarization photon is transferred to the electron through the
excitonic state, the exciton dephasing affects the state of the
electron. The exciton dephasing only reduces the fidelity by a few
percents due to the optical dephasing caused by population
relaxation or the loss of phase coherence among the dipoles and the
spin dephasing caused by spin interactions with the surrounding
nuclei in self-assemble In(Ga)As-based QDs. The optical coherence
time of excitons $T_2$ can be in several hundreds of picoseconds
range at low temperature while the cavity photon lifetime $\tau$ is
much shorter than the cavity photon lifetime 10 ps ($Q$:
$10^4-10^5$) \cite{opticldeph1,opticldeph2,opticldeph3}. The
coherence time of a QD-hole spin $T_2^h$ is longer than 100 ns
\cite{Hole-QD} and it is at least three orders of magnitude longer
than the cavity photon lifetime $\tau\sim10$ ps, so it can be
neglected. Besides the exciton dephasing, a few percent heavy-light
hold mixing in the valence reduces the fidelity by a few percents
\cite{opticselec}. This effect can be reduced by improving the
sample design and choosing different types of QDs.

\bigskip
\bigskip

{\large \textbf{Acknowledgments}}

  This work is supported by the National Natural Science Foundation of
China under Grant Nos. 11174039 and 11474026, and NECT-11-0031.

\bigskip

{\large \textbf{Authors contributions}}

H.R. and F.G.  wrote  the main  manuscript text  and  prepared
figures 1-3. H.R. prepared figures 4 and 5. F.G. supervised the whole project. Both authors reviewed
the manuscript.

\bigskip

{\large \textbf{Additional information}}

 Competing financial interests: The authors declare no competing
financial interests.

\end{document}